\documentclass[slac_one]{revtex4}
\usepackage{graphicx}
\usepackage{fancyhdr}
\usepackage{amsmath}
\newcommand{\PT}{\mathrm{PT}}
\newcommand{\NP}{\mathrm{NP}}
\newcommand{\as}{\alpha_s}
\newcommand{\VEV}[1]{\left\langle #1 \right\rangle}
\newcommand{\aeff}{\alpha_\mathrm{eff}}
\newcommand{\Kout}{K_\mathrm{out}}
\newcommand{\out}{\mathrm{out}}

\begin{document}

\title{Why multi-jet studies?}

%

\author{A. Banfi}
\affiliation{Universit\`a di Milano-Bicocca and INFN, Sezione di Milano, Italy}

\begin{abstract}
  We review the status of non-perturbative analyses of multi-jet event
  shape distributions and mean values, highlighting the physical
  insight on QCD dynamics they can provide. 
\end{abstract}

\maketitle

\thispagestyle{fancy}


\section{INTRODUCTION}
In every QCD observable, perturbative (PT) and non-perturbative (NP)
dynamics are inseparable, the reason being that the observed degrees
of freedom are not quarks and gluons, the elementary particles
entering the QCD bare Lagrangian, but hadrons, whose description in
terms of partons is well beyond the domain of PT theory.  Fortunately,
at least for sufficiently inclusive observables, the difference
between parton and hadron level predictions is suppressed by inverse
powers of the hard scale of the process. As far as the latter is
short-distance dominated, one can safely compute QCD observables using
the PT parton language, and interpret the discrepancy with
experimental data, in case any is seen, as the need for NP
hadronisation corrections. One can then adopt two complementary
approaches. One is to consider observables whose hadronisation
corrections are almost negligible, for instance total cross sections
or inclusive non-QCD particle distributions.  These observables can be
computed in PT QCD and exploited to determine the value of the
coupling $\as$~\cite{alpha-exp}. The other approach is to consider
observables which are very sensitive to NP physics, in order to have
an insight on the hadronisation mechanism. The best known example is
event shape variable distributions and mean values. These variables
are constructed by combining final state momenta to obtain a number
that gives an idea of the geometrical properties of hadron
energy-momentum flow.  The value of an event shape $V$ is related to
the scale at which hadrons are probed, so that measuring event shape
distributions makes it possible to study physics at very different
scales, which range from the domain of PT QCD ($V\sim 1$) down to the
confinement region ($V \sim \Lambda_\mathrm{QCD}$) where the
quark/gluon language is scarcely applicable. For this reason, although
shape variables have been used to measure $\as$, they are ideal for
investigating properties of QCD dynamics.  So far, both experimental
and theoretical investigations have been restricted only to two-jet
event shapes (see \cite{DSreview} for a review). Here we will discuss
how the extension of such studies to multi-jet event shapes can shed
further light on the interplay between PT and NP physics in QCD
observables.

\section{TWO-JET STUDIES}

\subsection{NP correction to event shapes}
\label{sec:NPcorr}
\begin{figure*}[h]
  \includegraphics[width=.5\textwidth]{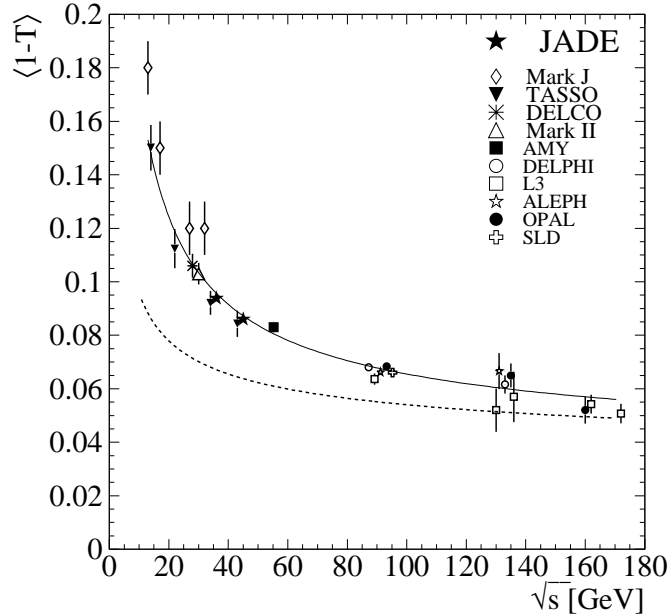}
  \caption{The mean value of $1\!-\!T$ as a function of $Q=\sqrt s$.
    Experimental data from various collaborations are plotted
    against theoretical predictions. The figure is taken from
    ref.~\cite{evshape-exp} \label{fig:meanT}.}
\end{figure*}

Experimental data clearly indicates that PT QCD alone is not
enough to predict event-shape distributions and mean
values~\cite{evshape-exp}.  Let us consider the mother of all event
shapes, the thrust~\cite{thrustdef}
\begin{equation}
  \label{eq:thrust}
  T = \max_{\vec n_T}\frac{ 
    \sum_h|\vec p_h \cdot \vec n_T|}{\sum_h |\vec p_h|}\>, 
  \qquad \tau \equiv 1-T\>. 
\end{equation}
The thrust is a measure of particle alignment, and is a typical
two-jet variable, since it vanishes in the limit of two narrow jets.
As one can see from fig.~\ref{fig:meanT}, the dependence of the mean
value of $1-T$ on the $e^+e^-$ centre-of-mass energy $Q$ is correctly
described only after adding to the QCD fixed order prediction a
NP $1/Q$-suppressed correction
\begin{equation}
  \label{eq:tau-mean}
  \VEV{\tau}=\VEV{\tau}_\PT + \VEV{\tau}_\NP\>,\qquad 
  \VEV{\tau}_\PT = \as(Q) \>\tau_1+\as^2(Q)\> \tau_2+\ldots \>,\qquad
  \VEV{\tau}_\NP \simeq \frac{C_\tau}{Q}\>,
\end{equation}
where $C_\tau\simeq 1\mathrm{GeV}$ when $\VEV{\tau}_\PT$ is evaluated
at next-to-leading order (NLO).

One might think that such a discrepancy could be removed by
including higher orders in the PT expansion. Actually Sterman
observed that a term $18 \>\alpha_s^3$ can already mimic a $1/Q$
behaviour~\cite{Sterman}.  However, theoretical analyses show that the
PT series is an intrinsically ill-defined object, since it is doomed
to diverge factorially (for a review, see~\cite{Beneke}).  Attempts to
regularise such a divergence give rise to a power-suppressed
ambiguity, known as infrared (IR) renormalon.  Looking more closely at
the origin of the divergence, one can see that it arises when
resumming {\it renormalon chain} graphs containing an arbitrary number
of linked quark or gluon bubbles~\cite{renormalon}. A renormalon
resummation for the thrust mean value yields
\begin{equation}
  \label{eq:tau-PT}
  \VEV{\tau}_{\PT} \simeq
  \frac{4C_F\as}{\pi}\sum_{n=0}^\infty  
  n!\left(\frac{2\beta_0\as}{4\pi}\right)^n n^{\beta_1}
  \simeq 2 C_F \int\frac{dk_t}{k_t}d\eta \>
  \frac{\as(k_t)}{\pi} \>\frac{k_t}{Q}e^{-|\eta|}\>,
\end{equation}
where $\beta_0$ and $\beta_1$ are the first two coefficients of the
QCD beta function. After a renormalon analysis, the series in
eq.~(\ref{eq:tau-PT}) gives a $1/Q$ ambiguity.  The last equality in
eq.~(\ref{eq:tau-PT}) results from the fact that the series can be
seen as the PT expansion of the integral of the running coupling
$\as(k_t)$ down to the infrared. Since the PT coupling develops a
Landau singularity at low momenta, one may be tempted to ascribe the
divergence to the presence of the Landau pole in the $k_t$ integration
contour.  However, eq.~(\ref{eq:tau-PT}) shows that the divergence of
$\VEV{\tau}_\PT$ is determined only by $\beta_0$ and $\beta_1$, i.e.
it is independent of the particular coupling adopted.  This naive
observation is confirmed by more refined theoretical analyses, which
show that IR renormalons are always present whatever is the behaviour
of the coupling at low scales~\cite{DU}.

The main message of this discussion is that, in order to obtain a
satisfactory {\it theoretical} understanding of event shape
observables, PT theory must always be supplemented with information on
QCD dynamics at low scales.  In particular, a size of the power
correction of around $1\mathrm{GeV}$ suggests that NP effects arise
from partons which have started the blanching process that leads
to the formation of hadrons.

\subsection{Two-jet shapes and the Feynman tube model}
\label{sec:Feynman}
Consider now the specific case of an event-shape $V$ in $e^+ e^-$
annihilation that vanishes in the two-jet limit (thrust,
$C$-parameter, heavy-jet mass, etc.). The mean value $\VEV{V}$ will
receive its main contribution from hard particles whose transverse
momenta (with respect to the thrust axis) are of the order of the hard
scale $Q$. There are however soft hadrons, with transverse momenta up
to about $1\mathrm{GeV}$, whose contribution to $\VEV{V}$ cannot be
fully computed with PT techniques.  The Feynman tube model~\cite{tube}
gives a phenomenological description of such hadrons by assuming that
their distribution is uniform in rapidity:
\begin{equation}
  \label{eq:dnh}
  \frac{dn_h}{d\ln k_t d\eta} = \Phi_h(k_t)\>.
\end{equation}
Observing also that their contribution $\delta V$ to $V$
is additive
\begin{equation}
  \label{eq:deltaV}
  \delta V \simeq \sum_i \frac{k_{ti}}{Q} f_V(\eta_i)\>,
\end{equation}
one obtains a $1/Q$-suppressed NP correction $\VEV{V}_\NP$ to 
$\VEV{V}$, given by
\begin{equation}
  \label{eq:VNP}
  \VEV{V}_\NP = \sum_h \int \frac{dk_t}{k_t}d\eta 
  \>\Phi_h(k_t) \frac{k_t}{Q} f_V(\eta) =
  c_V\frac{\VEV{k_t}_\NP}{Q}\>,\qquad   c_V = \int_{-\!\infty}^\infty\!\!\! d\eta 
\>f_V(\eta)\>,
\end{equation}
where $c_V$ is a calculable coefficient and
\begin{equation}
  \label{eq:ktnp}
  \VEV{k_t}_\NP = 
  \sum_h \int \frac{dk_t}{k_t} \Phi_h(k_t)\> k_t\>,
\end{equation}
is the average transverse momentum of the produced hadrons, and represents
a genuine NP quantity.

The assumption of eq.~(\ref{eq:dnh}) makes it possible to completely
factorise rapidity and transverse momentum dependence.  
Furthermore, eq.~(\ref{eq:VNP}) implies that
power corrections to two-jet event shapes are {\it universal}, in the
sense that, after computing the variable dependent coefficient $c_V$,
they depend only on the NP parameter $\VEV{k_t}_\NP$, which is the
same for all variables.

A comment is in order concerning the validity of eq.~(\ref{eq:dnh}).
Since in a hard collision most particles are aligned along the thrust
axis, one expects that the distribution of hadrons away from the jets
should be invariant over small boosts along the thrust axis direction,
but that, as soon as one moves forward in rapidity, eq.~(\ref{eq:dnh})
gets modified. Fortunately, for most event shapes $f_V(\eta)$ strongly
suppresses the contribution of hadrons at large rapidities, so that
leading power corrections are determined only by soft hadrons in a
central rapidity region, which are well described by
eq.~(\ref{eq:dnh}).\footnote{ Surprisingly enough, eq.~(\ref{eq:dnh})
  gives the correct leading power correction also for
  variables that are uniform in rapidity. This is due to a tricky
  interplay between PT and NP QCD radiation, which will be discussed
  in the next section.}

This very same approach can be used to compute power correction to
event-shape distributions. Since $V=V_\PT+\delta V$ with $\delta V
\sim \epsilon/Q$, the distribution $d\sigma/dV$ can be seen as a
convolution of a PT distribution $d\sigma_\PT/dV_\PT$ and a NP shape
function $f_\NP(\epsilon)$~\cite{shape-fun}
\begin{equation}
  \label{eq:dist-PT-NP}
  \frac{1}{\sigma}\frac{d\sigma}{dV} =
  \int dV_\PT\> d\epsilon \>\frac{1}{\sigma}\frac{d\sigma_\PT}{dV_\PT}(V_\PT) 
  \>f_\NP(\epsilon)\> \delta\left(V-V_\PT-\frac{\epsilon}{Q}\right)\>.
\end{equation}
In the region $VQ \gg \VEV{\epsilon}$ one can expand $f_\NP(\epsilon)$ around
$\VEV{\epsilon}$ and obtain 
\begin{equation}
  \frac{1}{\sigma}\frac{d\sigma}{dV}
  \simeq
  \frac{1}{\sigma}\frac{d\sigma_\PT}{dV}
  \left(V-\frac{\VEV{\epsilon}}{Q}\right)\>.
\end{equation}
Recalling now that $\VEV{\epsilon}\!/Q = \VEV{V}_\NP$, one observes that
leading power corrections to $d\sigma/dV$ result simply in a shift of
the corresponding PT distribution, whose magnitude is the same as the
power correction to $\VEV{V}$~\cite{DW}.

To conclude, the basic assumption of the Feynman tube model, that the
distribution of soft (central) hadrons is uniform in rapidity, implies
that leading $1/Q$ NP corrections both to mean values and
distributions can be expressed in terms of a single universal
parameter $\VEV{k_t}_\NP$. This universality hypothesis has been
tested against experimental data and is found to hold within $20\%$
accuracy~\cite{DSreview}.

\section{EXTENSION TO MULTI-JET EVENT SHAPES}

\subsection{Intra-jet hadron distribution}
\label{sec:intra-jet}
When moving from two-jet to multi-jet events one encounters
difficulties in extending eq.~(\ref{eq:dnh}). One can reasonably think
that particles inside one of the jets (inter-jet hadrons) are produced
uniformly in rapidity and azimuth with respect to the jet axis.
However, for particles at large angles with respect to all jets
(intra-jet hadrons), since there is no natural way to define $\eta$ and
$\phi$, one does not expect $dn_h$ to have a simple
expression.  To find a reasonable assumption for $dn_h$, we start from
two-jet events and consider the PT probability $dw(k)$ for the
emission of a soft dressed gluon $k$ off a back-to-back
quark-antiquark system (whose momenta are $p$ and $\bar p$) in a
colour singlet~\cite{thrust-np}:
\begin{equation}
  \label{eq:dw-2jet}
  dw(k) = C_F \frac{dk_t^2}{k_t^2} d\eta \frac{d\phi}{2\pi}
  \frac{\as(k_t)}{\pi}\>,\qquad
  \eta = \frac 12\ln\frac{\bar p k}{pk}\>,\qquad
  k_t^2 = \frac{(2pk)(2k \bar p)}{2p\bar p}\>,
\end{equation}
where $\eta$ and $k_t$, the gluon rapidity and transverse momentum,
have been written in a Lorentz invariant form and the coupling is
taken in the physical CMW scheme~\cite{CMW}.  The inter-jet hadron
distribution in eq.~(\ref{eq:dnh}) can be obtained from
eq.~(\ref{eq:dw-2jet}) by replacing $2\>C_F \as(k_t)/\pi$ with $\sum_h
\Phi_h(k_t)$. In this way $dn_h$ might be interpreted as an effective
measure of QCD interaction strength in the infrared. When the number
of hard emitters is larger than two, the distribution $dw(k)$ can be
written as a sum of contributions from all dipoles formed by the hard
partons~\cite{CatGraz}:
\begin{equation}
  \label{eq:dw-multi}
  dw(k) = \sum_{i<j}(-\vec T_i\cdot \vec T_j) 
  \frac{d\kappa^2_{ij}}{\kappa^2_{ij}} d\eta_{ij} \frac{d\phi_{ij}}{2\pi}
  \frac{\as(\kappa_{ij})}{\pi}\>,\qquad
  \eta_{ij} = \frac 12 \ln\frac{p_j k}{p_i k}\>, \qquad
  \kappa_{ij}^2 = \frac{(2p_i k)(2k p_j)}{2p_i p_j}\>,
\end{equation}
where now $\kappa_{ij}$, $\eta_{ij}$ and $\phi_{ij}$ are the
transverse momentum, rapidity and azimuth in the $(ij)$-dipole
centre-of-mass frame, and the coupling is again in the CMW scheme. The
vector $\vec T_i$ represents the colour charge of hard
parton $p_i$. From colour conservation one has that for less than four
emitters $\vec T_i\cdot \vec T_j$ are numbers, while
starting from four partons they are actual matrices in colour space.

We now postulate that, after a suitable extension of the CMW coupling
at low scales has been introduced, eq.~(\ref{eq:dw-multi}) gives the
distribution of intra-jet hadrons in a multi-jet event. This
corresponds to the so-called local parton-hadron duality (LPHD)
hypothesis (see \cite{LPHD} for a recent review), which states that
hadron flow is determined by parton flow. This assumption has solid
phenomenological bases, since it describes very well hadron
multiplicity flows~\cite{multi} and string/drag effects in three-jet
events~\cite{drag}.

Before constructing such an extended coupling, we first discuss to
what extent the soft large-angle approximation is sufficient to
compute $1/Q$ power corrections to event shapes. In fact, $\kappa_{ij}$ can
reach the NP domain not only for soft large-angle, but also for hard
collinear emissions. Considering emissions collinear to a given leg,
we can classify event shapes according to whether they are damped in
rapidity or not.
 
If $V$ is damped in rapidity, soft collinear particles with transverse
momentum $k_{ti}$ and rapidity $\eta_i$ with respect to the given leg
contribute to $V$ with a correction
\begin{equation}
  \label{eq:deltaV-damped}
  \delta V \simeq \sum_i \frac{k_{ti}}{Q} f_V(\eta_i)\>,\qquad
  f_V(\eta) \sim e^{-\eta} \>, \quad\>\eta\to\infty\>.
\end{equation}
This implies that only large angle emissions contribute to $\delta V$,
up to higher power corrections.

If $V$ is uniform in rapidity, we have $\delta V \sim \sum_i k_{ti}$,
so that one might think that all rapidities contribute equally to
$\delta V$. However, due to PT radiation, each hard parton takes a
recoil, and acquires a transverse momentum $p_t \sim VQ$. But one
measures transverse momenta with respect to a fixed axis (e.g. the
thrust axis), and not with respect to the hard emitter's direction
(the jet broadenings~\cite{broadNP} are a relevant example). Since, as
we have seen in the previous section, leading power corrections come
from the region $\delta V \ll V$, we must have then $k_{ti}/Q \sim
\delta V \ll p_t/Q \sim V$.  This means that particles giving the
leading contribution to $\delta V$ must be soft, and displaced from
the hard emitting parton, i.e.  soft at large angles.  To conclude,
eq.~(\ref{eq:dw-multi}) gives the physically correct starting point to
compute $1/Q$ power corrections to event-shape variables.

\subsection{Dokshitzer-Marchesini-Webber (DMW) extension of CMW
  coupling}
\label{sec:DMW}

The CMW coupling can be extended at low scales via the dispersion
relation~\cite{DMW}
\begin{equation}
  \label{eq:DMW}
  \as(\kappa) = -\int_0^\infty\!\!\frac{dm^2}{(m^2+\kappa^2)}\>\rho_s(m^2)=
  \kappa^2 \int_0^\infty\!\!\!\frac{dm^2}{(m^2+\kappa^2)^2}\>\aeff(m^2)\>,
\end{equation}
where $\aeff(m^2)$ is the logarithmic derivative of the spectral
density $\rho_s(m^2)$, and is itself a QCD coupling.

This approach automatically incorporates the LPHD hypothesis by
assuming that hadronisation corrections are due to {\it gluers},
extra-soft gluons with transverse momenta $\kappa \sim
\Lambda_\mathrm{QCD}$, whose emission probability is ruled by the NP
part of the dispersive coupling.

Within the dispersive DMW approach, power corrections to event shapes can be
computed by considering the emission of a single gluer, and following
a standard procedure~\cite{np-ee}:
\begin{enumerate}
\item real and virtual corrections are combined to obtain the dispersive
  representation of the QCD coupling. This means that in
  eq.~(\ref{eq:dw-multi}), for each dipole, one can make the substitution
  \begin{equation}
    \label{eq:dw-DMW}
    \frac{\as(\kappa)}{\kappa^2} = 
    \int_0^\infty\!\!\!\frac{dm^2}{(m^2+\kappa^2)^2}\aeff(m^2)\>;
  \end{equation}
\item the gluer is given a mass $m^2$, i.e. it is allowed to decay
  inclusively. In practice this results in replacing everywhere in the
  event-shape definition $\kappa^2$ with $\kappa^2+m^2$, so that the
  contribution to a variable $\delta V^{(ij)}_\mathrm{gluer}$ of a
  gluer emitted off the $(ij)$-dipole is
  \begin{equation}
    \label{eq:dV-gluer}
    \delta V_\mathrm{gluer}^{(ij)} = \frac{\sqrt{\kappa^2+m^2}}{Q} 
    f_V^{(ij)}(\eta,\phi)\>,
  \end{equation}
  where $\kappa$, $\eta$ and $\phi$ are, as usual, considered in the
  emitting dipole centre-of-mass frame.  Then the leading power
  correction to $\VEV{V}$ becomes
  \begin{equation}
    \label{eq:dV-naive}
    \VEV{V}^\mathrm{naive}_\NP = 
    \frac{\VEV{\kappa}_\mathrm{naive}}{Q} 
    \sum_{i<j} (-\vec T_i \cdot \vec T_j)\> c_V^{(ij)}\>, \qquad
    c_V^{(ij)} = \int d\eta \frac{d\phi}{2\pi} f_V^{(ij)}(\eta,\phi)\>,
  \end{equation}
  where 
  \begin{equation}
    \label{eq:kappa-naive}
    \VEV{\kappa}_\mathrm{naive} = \int_0^\infty 
    \frac{d\kappa^2 dm^2}{(m^2+\kappa^2)^2} \sqrt{\kappa^2+m^2}
    \>\frac{\aeff(m^2)}{\pi} \simeq 
    \frac{2}{\pi} \int_0^\infty \frac{dm^2}{m^2} m \>\delta \aeff(m^2)\>,
  \end{equation}
  turns out to be related to the $1/2$-moment of $\delta \aeff(m^2)$,
  the NP part of the dispersive coupling;
\item non-inclusiveness of event shapes with respect to secondary
  gluon decay is accounted for by replacing
  $\VEV{\kappa}_\mathrm{naive}$ with $\VEV{\kappa}_\NP\equiv {\cal
    M}\VEV{\kappa}_\mathrm{naive}$, where ${\cal M}$ is the so-called
  {\it Milan} factor~\cite{thrust-np,Milan}. It is also customary to
  rewrite $\VEV{\kappa}_\NP$ in terms of $\alpha_0(\mu_I)$, the
  average of the dispersive coupling below the merging scale $\mu_I$,
  in such a way that the sum of PT and NP contributions to $\VEV{V}$
  is free of IR renormalons:
  \begin{equation}
    \label{eq:alpha0}
    \VEV{\kappa}_\NP = \frac{4 \mu_I}{\pi^2}{\cal M}
    \left(\alpha_0(\mu_I)-\as(Q)+{\cal O}(\as^2)\right)\>, \qquad
    \alpha_0(\mu_I) = \int_0^{\mu_I}\frac{dk}{\mu_I}\as(k)\>.
  \end{equation}
\end{enumerate}

We are now able to discuss what kind of information on QCD dynamics we
obtain from power corrections to two- or multi-jet event shapes. 

From eq.~(\ref{eq:dV-naive}) the power correction to a two-jet
variable can be written in the form
\begin{equation}
  \label{eq:dV-2jet}
  \VEV{V}_\NP = C_F \frac{\VEV{\kappa}_\NP}{Q} c_V = 
  \frac{\VEV{k_t}_\NP}{Q} c_V\>, 
\end{equation}
where $\VEV{k_t}_\NP$ is the NP parameter appearing in
eq.~(\ref{eq:ktnp}), and the last equality has been obtained by
comparing eq.~(\ref{eq:VNP}) and eq.~(\ref{eq:dV-naive}). We see then
that both the dispersive approach and the Feynman tube model give the
same result for $\VEV{V}_\NP$. In fact universality of leading NP
corrections to two-jet event shapes implies simply that central soft
hadrons are produced uniformly in rapidity.

When increasing the number of jets, the NP correction to a multi-jet
variable is given by
\begin{equation}
  \label{eq:dV-multi}
    \VEV{V}_\NP=\frac{\VEV{\kappa}_\NP}{Q} \>\tilde c_V\>, \qquad
    \tilde c_V\equiv\sum_{i<j} (-\vec T_i \cdot \vec T_j) \> c_V^{(ij)}\>.
\end{equation}
The main features of this PT QCD inspired correction are:
\begin{enumerate}
\item the same parameter $\VEV{\kappa}_\NP$ determines the magnitude
  of power corrections for {\it all} event shapes and for {\it any
    number} of hard jets;
\item $\VEV{V}_\NP$ depends on the {\it colour} charges of emitting
  partons through the factors $\vec T_i \cdot \vec T_j$;
\item the power correction is sensitive to the {\it geometry} of the
  underlying hard event (the angles between the jets) through the
  coefficients $c_V^{(ij)}$.
\end{enumerate}
This colour and geometry dependence is a highly non-trivial property
that, once established, would imply that hadronisation preserves the
main characteristics of partonic energy-momentum flow.

\subsection{Three-jet event shapes in $e^+e^-$ annihilation}
\label{sec:3jet-ee}
We now consider the specific case of three-jet events in $e^+e^-$
annihilation. At Born level a three-jet event is made up of a quark
$p_q$, an antiquark $p_{\bar q}$ and a gluon $p_g$. Transverse momenta
are, as usual, defined with respect to the thrust axis $\vec n_T$, and
one defines an event-plane as the one containing $\vec n_T$ and thrust
major axis $\vec n_M$, defined as the one maximising the projection of
transverse momenta. Every transverse momentum $\vec k_t$ can be
decomposed into an in-plane component $k^\mathrm{in}$ and an
out-of-plane component $k^\out$ as follows
\begin{equation}
  \label{eq:in-out}
 \vec k_t = k^\mathrm{in} \>\vec n_\mathrm{in} + k^\out \>\vec n_\out  \>,\qquad
 \vec n_\mathrm{in}=\vec n_M\>, \quad
 \vec n_\out \equiv \vec n_T\times\vec n_M\>. 
\end{equation}
The two variables for which there exists a prediction for leading
power corrections are the $D$-parameter~\cite{dpar} and the
thrust-minor $T_m$ (a.k.a. $\Kout$)~\cite{kout-ee}.

The $D$-parameter is defined as the determinant of the tensor
\begin{equation}
  \label{eq:Dpar}
  \theta^{\alpha\beta} Q \equiv \sum_h \frac{p_h^\alpha p_h^\beta}{|\vec p_h|}\>, 
  \qquad D \equiv 27 \det \theta = 27 \lambda_1 \lambda_2 \lambda_3 \>.
\end{equation}
The soft particle contribution $\delta D$ is given by
\begin{equation}
  \label{eq:deltaD}
  \delta D \simeq 27 \lambda_1 \lambda_2 
  \sum_i \frac{\kappa_i^2 \sin^2 \phi_i}{\omega_i Q} \>,
\end{equation}
where $\kappa_i$ is the transverse momentum in the emitting dipole
centre-of-mass frame and $\phi_i=0$ for an emission inside the event
plane.  Due to the presence of the energy $\omega_i$ in each
denominator of eq.~(\ref{eq:deltaD}), this variable is damped in
rapidity, and the coefficient $\tilde c_D$ is given by
\begin{equation}
  \label{eq:cD}
  \tilde c_D = C_F c_D^q(T,T_M) + C_F c_D^{\bar q}(T,T_M) + C_A c_D^g(T,T_M)\>,
\end{equation}
where $c_D^q$, $c_D^{\bar q}$ and $c_D^g$ depend on the event geometry
through $T$ and $T_M$.

The thrust-minor $T_m$ measures the out-of-event-plane momentum flow:
\begin{equation}
  \label{eq:Tm}
  T_m Q = \Kout \equiv \sum_h |p_h^\out| \>,\qquad
  \delta\Kout \simeq \sum_i \kappa_i |\sin \phi_i| \>.
\end{equation}
This variable is uniform in rapidity, so that one needs to consider
the fact that each hard emitting parton recoils against PT radiation.
This results in
\begin{equation}
  \label{eq:ckout}
  \tilde c_{\Kout} \simeq \VEV{|\sin\phi|}\left(C_F\ln\frac{Q_{q\bar q}}{|p_q^\out|}+
    C_F\ln\frac{Q_{q\bar q}}{|p_{\bar q}^\out|}+
    C_A\ln\frac{Q_{qg} Q_{g\bar q}}{Q_{q\bar q} |p_g^\out| }\right)\>,\qquad
  \VEV{|\sin\phi|}\equiv\int_0^{2\pi}\frac{d\phi}{2\pi}|\sin \phi|=\frac 2\pi \>, 
\end{equation}
where $Q^2_{ij}=2 p_j \cdot p_j$. We notice that for each hard emitter
the power correction is proportional to its colour charge and to the
rapidity interval available to the emitted gluon. The fact that the NP
correction depends on hard parton recoil implies that the final answer
will be obtained after averaging $\tilde c_{\Kout}$ over the PT recoil
distribution, which results in a complicated but very interesting
interplay between PT and NP physics.

\begin{figure}[h]
  \includegraphics[width=.45\textwidth,height=.3\textheight]
  {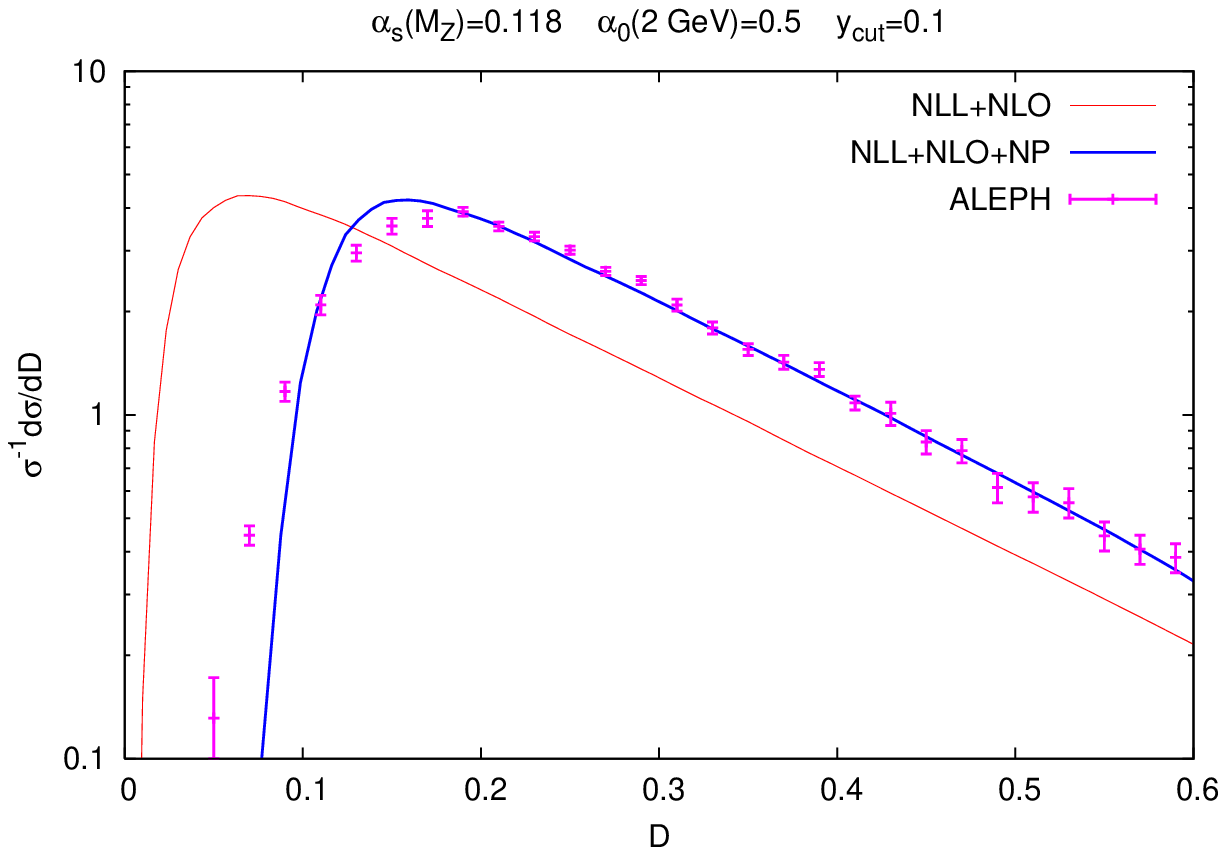}
  \includegraphics[width=.45\textwidth,height=.3\textheight]
  {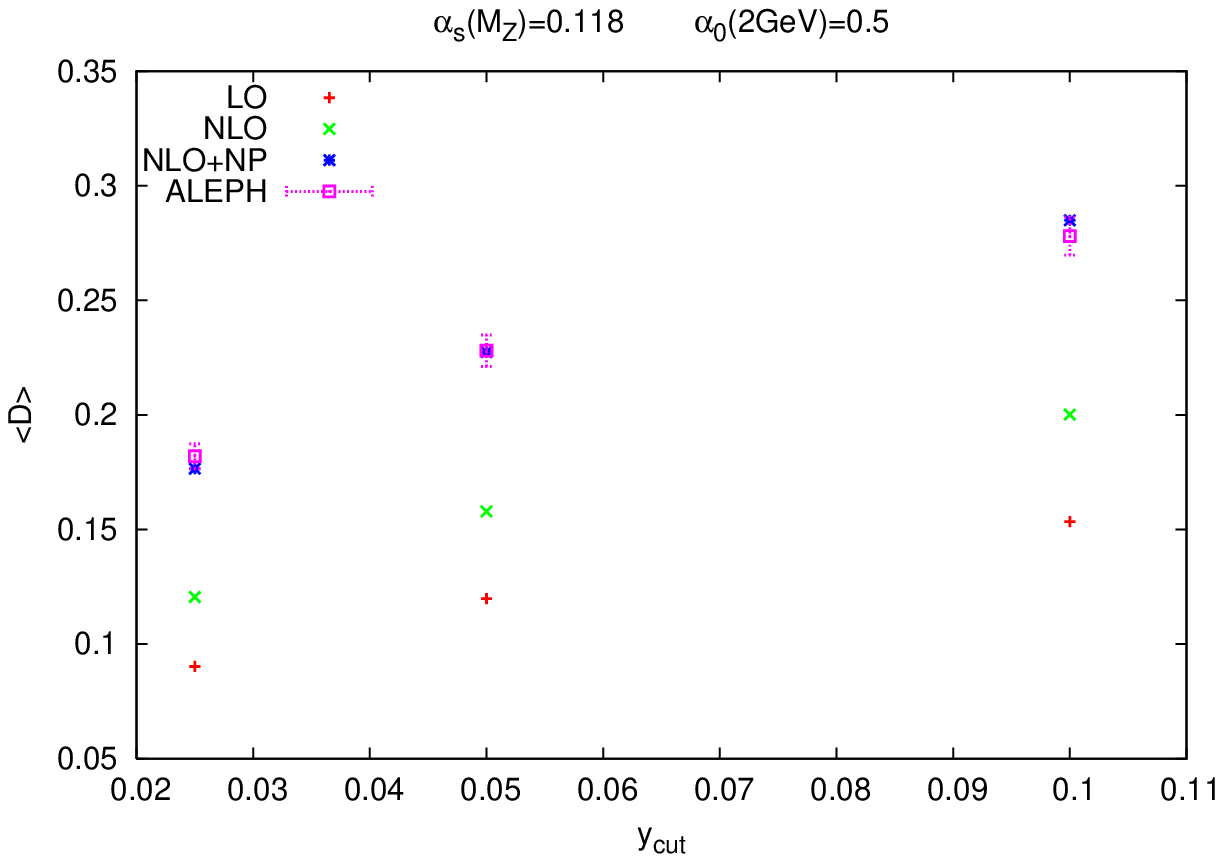}
  \caption{Theoretical predictions~\cite{dpar} for distribution (left)
    and mean values (right) of the $D$-parameter compared to ALEPH
    data~\cite{aleph-data}, for $Q=91.2\mathrm{GeV}$. Mean values for
    the data are computed from the corresponding distributions, with a
    $2\%$ estimated experimental error and an extra $1\%$ systematic
    error to account for the extraction of mean values from
    distributions~\cite{hasko}.}
  \label{fig:Dpar}
\end{figure}

Fig.~\ref{fig:Dpar} shows the comparison between theoretical
predictions~\cite{dpar} and experimental data~\cite{aleph-data} for
the $D$-parameter differential distribution and mean values. Three-jet
events are selected by imposing a lower limit $y_\mathrm{cut}\!=\!0.1$ on the
(Durham) three-jet resolution $y_3$. The distribution in
fig.~\ref{fig:Dpar} represents the first theoretical prediction for a
three-jet variable at the state-of-the-art accuracy, which includes
all-order PT resummation of logarithmic enhanced contribution to
next-to-leading logarithmic (NLL) accuracy, matching to the corresponding
NLO distribution obtained with \textsc{nlojet}++~\cite{nlojet}, and
$1/Q$ NP power corrections. For each Born configuration, the leading
NP correction results in a shift of the PT distribution. The presented
distribution is also slightly squeezed since the shift depends on the
event geometry and one has to integrate over different Born
configurations.
The comparison to data is overall quite good. One can observe that the
shift is quite large, about twice as much as in two-jet event shapes.
This is mainly due to NP radiation off a gluon, which has a large
colour charge. Although a full analysis of theoretical uncertainties
remains to be performed~\cite{AG-hardwork}, the magnitude of the shift
suggests that leading power corrections might not describe with
sufficient accuracy the distribution around the peak, but that higher
powers, or even a shape function, might be needed.

\subsection{Three-jet event shapes in hadron-hadron collisions}
\label{sec:3jet-hh}
One can construct three-jet event shapes also in hadron-hadron
collisions. The simplest case is to consider a process with an
electro-weak vector boson $Z^0$ or $W^\pm$ recoiling against a
high-$p_t$ jet. The process involves three hard emitters, two
incoming ($p_1$ and $p_2$) and one outgoing ($p_3$).

The event plane is the one containing the beam and the momentum $\vec
q$ of the vector boson, so that we have $\vec n_\mathrm{out} = (\vec q
\times \vec p_1)/(|\vec q||\vec p_1|)$. Since one does not measure
inside the beam pipe, any measured hadron $p_h$ will have $|\eta_h| <
\eta_0$, where $\eta_0$ is the maximum available rapidity. Therefore
the out-of-plane momentum flow $\Kout$ has to be defined
as~\cite{kout-hh}
\begin{equation}
  \label{eq:kout-hh}
  \Kout \equiv \sum_h |p_h^\out| \times \Theta(\eta_0-|\eta_h|)\>,
\end{equation}
In the soft limit
$\delta\Kout$ gets two contributions, one from particles emitted
directly from the three hard partons, the other from hadrons produced
in beam remnant interactions, the so-called soft underlying
event~\cite{pedestal}:
\begin{equation}
  \label{eq:deltakout-hh}
  \delta\Kout = \delta\Kout^\mathrm{gluer}+\delta\Kout^\mathrm{remnant}\>,\qquad
  \delta\Kout^\mathrm{gluer} \simeq {\sum_i}^\prime \kappa_i |\sin\phi_i| \>,
  \qquad 
  \delta\Kout^\mathrm{remnant} \simeq{\sum_i}^\prime k_{ti} |\sin\phi_i|\>,
\end{equation}
where each primed sum is restricted to particles outside the beam pipe.
In the term $\delta\Kout^\mathrm{gluer}$ the transverse momentum
$\kappa_i$ is considered as usual in the emitting dipole
centre-of-mass frame, while in $\delta\Kout^\mathrm{remnant}$ each
$k_{ti}$ is in the laboratory frame.

As before, the gluer contribution depends on the same parameter
$\VEV{\kappa}_\NP$ as for $e^+e^-$ event shapes, and is given by:
\begin{equation}
  \label{eq:deltakout-gluer}
  \delta\Kout^\mathrm{gluer} = \VEV{\kappa}_\NP\> \tilde c_{\Kout} \>,
  \qquad
  \tilde c_{\Kout} \simeq \VEV{|\sin\phi|} 
  \left(C_1(\eta_0-\eta_3)+C_2(\eta_0+\eta_3)+
    C_3 \ln\frac{Q_t}{|p_3^\out|}\right)\>,
\end{equation}
where $Q_t$ is the transverse momentum of the vector boson, which
represents also the hard scale of the process, and $C_i$ is the colour
charge of hard parton $p_i$. As in the $e^+e^-$ case, for each
emitting hard parton, the power correction is proportional to the
rapidity interval available to the emitted gluer.

The beam remnant contribution can be computed by assuming that the
distribution of hadrons produced in soft beam-remnant interactions is
uniform in rapidity, which gives
\begin{equation}
  \label{eq:deltakout-remnant}
  \delta\Kout^\mathrm{remnant} = 
  \VEV{|\sin\phi|}\times\VEV{k_t}_\mathrm{remnant} 
  \times 2\eta_0\>. 
\end{equation}
We have then access to the new NP parameter
$\VEV{k_t}_\mathrm{remnant}$, the average transverse momentum of hadrons
produced in beam remnant interactions. This very same parameter will
appear in power corrections to event shapes in hadronic dijet
production~\cite{hh-shapes}.

\section{A RICH PHENOMENOLOGICAL PROGRAM}
\label{sec:pheno}
The considerations we have presented so far can be seen as the
starting point for a rich phenomenological program.  First of all
one has to keep in mind that multi-jet event-shape observables are not
ideal to perform pure PT studies aimed at precise determinations of
the strong coupling $\as$. Large theoretical uncertainties are
associated to PT higher orders and to NP power corrections, and jet
selection cuts reduce the number of available events. However, from
the point of view of understanding QCD dynamics, they are an
incredibly valuable tool. A research program has already started with
the study of power corrections to three-jet event shapes in $e^+e^-$
annihilation and DIS~\cite{AG-hardwork}, aimed at determining whether
the NP parameter $\alpha_0$ is the same as for two-jet event shapes.
Once the universality of $\alpha_0$ is established, one could move to
considering hadron-hadron collisions. There one could test our
understanding of beam-remnant interactions in a complementary way with
respect to what has been done so far. At the moment the only
observable used to test models of the underlying event is the
away-from-jet particle/energy flow~\cite{Field}, for which we have a
less clear theoretical understanding than we have for event shapes.
Provided our modelling of the underlying event proves to be correct,
we have access to the NP parameter $\VEV{k_t}_\mathrm{remnant}$, and
are able to use this parameter to study event-shapes in hadronic dijet
production~\cite{hh-shapes}, where a lot of progress has been done in
recent years, especially since when the automated resummation program
\textsc{caesar} has become available~\cite{caesar}.

\begin{acknowledgments}
  It was Yuri Dokshitzer and Pino Marchesini who initiated me to
  multi-jet studies. I therefore thank them and Giulia Zanderighi for
  a many-year collaboration on this subject. I am also grateful to the
  organisers for the possibility to participate to this very
  interesting and stimulating workshop. Finally, I wish to thank Gavin
  Salam for discussions and suggestions on the phenomenological analysis of
  the $D$-parameter.
\end{acknowledgments}

\vspace{-.5cm}


\end{document}